\documentclass[runningheads]{svmult}

\usepackage{makeidx}   
\usepackage{graphicx}  
\usepackage{subeqnar}  
\usepackage{multicol}  
\usepackage{physprbb}  
\makeindex             

\font\tbs=cmbsy10
\def\bnabla{\hbox{\tbs\char'162}}

\begin{document}
\title*{Super-Eddington black hole accretion:}
\subtitle{Polish doughnuts and slim disks}
%
%
\titlerunning{Super-Eddington accretion}
%
\author{Marek A. Abramowicz$^{*\dag}$}

\authorrunning{M.A. Abramowicz}
%
%
\institute{Nordita, Copenhagen, Denmark
\\~~
\\$^*$On leave from: Faculty of Natural Sciences, Theoretical Physics, G\"oteborg University
\\$^{\dag}$Address: Chalmers University, S-412-96 G\"oteborg, Sweden, {\it marek@fy.chalmers.se}}

\maketitle              

\begin{abstract}
The theory of highly super-Eddington black hole accretion, ${\dot M} \gg {\dot M}_{\rm Edd}$, was developed in the 1980s in Warsaw by Paczy\'nski and his collaborators \cite{AbrJarSik80}--\cite{PacAbr...82} in terms of {\it Polish doughnuts}$^1$, i.e. low viscosity, rotating accretion flows that are optically thick, radiation pressure supported, cooled by advection, and radiatively very inefficient. Polish doughnuts resemble fat tori with two narrow funnels along rotation axis. The funnels collimate radiation into beams with highly super-Eddington luminosities. {\it Slim disks}\footnote{The name {\it Polish doughnuts} was coined by M.Rees, and {\it slim disks} by A.Ko{\l}akowska.} introduced later by Abramowicz, Lasota and collaborators first in \cite{AbrLasXuC86}, and more fully in \cite{AbrCzeLas88} have only moderately super-Eddington accretion rates, ${\dot M} \ge {\dot M}_{\rm Edd}$, rather disk-like shapes, and almost thermal spectra \cite{SzuMalAbr96}-\cite{BlaHubKro01}.  
\end{abstract}

\section{Introduction}

Is there an upper limit to the accretion rate ${\dot M}$, preventing astrophysical black holes to swallow all the matter infalling into them? Many astrophysicists are convinced that the answer to this fundamental question should be: {\it yes, the limit exist and it is given by the standard Eddington accretion rate},

\begin{equation}
{\dot M}_{\rm Edd} = {L_{\rm Edd} \over c^2}  =1.5 \times 10^{17} \left ({M \over M_{\odot}}\right)  [\,{{\rm g}\,\,{\rm sec}^{-1}}].
\end{equation}

\noindent However, because there is a clear observational evidence for super-Eddington luminosities, e.g. from X-ray binaries (SS433, GRS 1915+105), and because super-Eddin\-gton luminosities powered by accretion imply highly super-Eddington accretion rates, this almost unanimous consensus\footnote{All works on cosmic structure formation reported at this Conference, used ${\dot M}_{\rm Edd}$ as the upper limit for the black hole growth rate.} is to me rather befuddling. 

I review here fundamental theoretical predictions about super-Eddington accretion: (1)$\,$super-Eddington luminosities are typical for rotating, radiation pressure supported black hole accretion flows, (2)$\,$these flows have very small efficiency and therefore they must have highly super-Eddington accretion rates, (3)$\,$super-Edd\-in\-gton accretion does not necessarily imply strong outflows. 

These predictions are solidly based on standard physics, and it should not be surprising that recent 3D hydro and MHD supercomputer simulations provide many beautiful illustrations for them \cite{IguCheAbr96}-\cite{IguNarAbr03}.  

\subsection{Sub-Eddington accretion: standard thin disks and adafs}

The sub-Eddington accretion is far more familiar to astrophysicists. Well known theory predicts for ${\dot M} \ll {\dot M}_{\rm Edd}$ two types of stable accretion, depending on the optical depth, $\tau $. 

When $\tau \gg 1$, accretion is described in terms of the famous {\it standard thin disk} model by Shakura \& Sunyaev \cite{ShaSun...74}. Standard thin disks are supported by gas pressure, cooled by radiation, and very efficient. They are geometrically thin in the vertical direction, $H/R \ll 1$. 

When $\tau \ll 1$, accretion is described by the {\it adaf}$\,$\footnote{The name {\it adaf} was coined by J.-P.Lasota.} model, propheted in 1977 by Ichimaru in a paper \cite{Ich......77} that was ignored by everybody for twenty years\footnote{However, many elements of the adafs theory were present in the influential ``ion tori'' paper by Rees, Phinney, Begelman \& Blandford \cite{BlaBegPhi82}.}. Only after their rediscovery in the mid 1990 by Narayan\footnote{R.Narayan did more than anybody else to develop the adaf idea into a useful and mature astrophysical theory.} \& Yi \cite{NarYi....94}, \cite{NarYi....95}, and by Abramowicz, Chen, Kato, Lasota \& Regev \cite{AbrCheKat95}, adafs started to be intensely studied by many authors (see reviews in \cite{Las.....a99}, \cite{Las.....b99}). Adafs are cooled by advection. They are very inefficient, geometrically thick, $H/R \sim 1$, and very hot (close to the virial temperature). Because of their very low efficiency, they are much less luminous than the standard thin disks. 

\section{Fundamental limits for power: Planck and Eddington}

The Planck power (i.e. power expressed in Planck's units) equals 

\begin{equation}
L_{\rm Planck} = {c^5 \over G} \approx 10^{58}~{\rm [erg\,sec^{-1}]} = 10^{52}~{\rm [Watts]}. 
\end{equation}

\noindent Rather surprisingly, it does not depend on the Planck constant $h$. The maximal energy available from an object with the mass $M$ (and gravitational radius  $R_{\rm G}=GM/c^2$) is $E_{\rm max} = Mc^2$. The minimal time in which this energy may be liberated is $t_{\rm min} =R_{\rm G}/c$. Thus, the maximal power $L_{\rm max} = E_{\rm max}/t_{\rm min} = c^5/G = L_{\rm Planck}$. This  is the absolute upper limit for power of anything in the Universe: all objects, phenomena, explosions, and evil empires\footnote{A power needed for the Creation was the rate at which the Big Bang transferred energy from a pre-Planck to the post-Planck state. For the reason outlined here, no more than $10^{52}~{\rm Watts}$ was needed to create the Universe.}. 

For example, a sphere with radius $R$ containing blackbody radiation at temperature $T$ radiates power $L \sim (caT^4) R^2$. The gravitational mass of the radiation inside the sphere is $M \sim (aT^4/c^2)R^3$, where $a = 8\pi^5k^4/15c^3h^3$ is the radiation constant. Thus $R_{\rm G} \sim (aT^4G/c^4)R^3$, and from $R_{\rm G} < R$ one gets $L < c^5/G$. 

Consider a stationary object with mass $M$ in which gravity and radiation are in equilibrium. Sikora \cite{Sik......80} noticed that the upper limit for the object radiative power may be expressed by $L_{\rm Edd} = L_{\rm Planck} {\Sigma_{\rm grav} / \Sigma_{\rm rad} }$. Here $\Sigma_{\rm grav}$ is the object total effective gravity cross section, and $\Sigma_{\rm rad}$ is its total radiative cross section. Sikora's expression is the most general version of the Eddington limit. Eddington himself, first in \cite{Edd......16}, and then in \cite{Edd.....a17}-\cite{Edd......26}, considered a much more specific case of a radiation pressure supported star\footnote{In 1913, shortly before the war, Bia{\l}obrzeski
\cite{Bia......13} pointed out that radiation pressure may be important in
stellar equilibria (see also \cite{Bia......16}-\cite{Bia......31}).
Eddington independently discovered the same three years later
\cite{Edd......16}, already during the war. Eddington, as a Quaker,
avoided an active war service and was able to continue his research during
the war years 1914-1918 at Cambridge, as the Director of the Observatory
there. Only after the war ended, the two scientists could communicate, and
Eddington wrote to Bia{\l}obrzeski: {\it I congratulate you on having been
apparently the first to point out the large share of radiation pressure in
internal equilibrium of stars} (quotation after \cite{Act......54}).
Bia{\l}obrzeski's contribution (and priority) was remembered by
astronomers still in the 1930s, (see the 1936 monograph by Tiercy
\cite{Tie......35} and the 1939 book by Chandrasekhar \cite{Cha......39}),
but today is almost totally forgotten~--- Chandrasekhar \cite{Cha......83}
makes no mention of it in his 1983 book on Eddington. (A.K.~Wr{\'o}blewski
\cite{Wro......04} provided most of historical information for this
footnote.)}, assuming that radiation interacts with matter by electron scattering (thus $\Sigma_{\rm rad} = N\sigma_T$ in Sikora's expression), and that effective gravity is provided by the Newton gravity alone (thus $\Sigma_{\rm grav} = 4 \pi R_G^2$). Here $N$ is the number of electrons in the object ($N=M/m_{\rm P}$ for pure hydrogen, $m_{\rm P}$ is the proton mass), $\sigma_T = 8\pi e^4/ 3m_{\rm e}^2 c^2$ is the Thomson cross section ($e$ is the electron charge and $m_{\rm e}$ its mass). In this specific case, Sikora's elegant argument immediately gives the standard formula for the Eddington limit, 

\begin{equation}
L_{\rm Edd} = L_{\rm Planck} {\Sigma_{\rm grav}\over \Sigma_{\rm rad} } =
{{4 \pi G M m_{\rm P} c} \over \sigma_{\rm T}}
= 1.4 \times 10^{38} \left ({M \over M_{\odot}}\right)  [\,{\rm erg}\,\,{\rm sec}^{-1}]\,.
\end{equation}

\noindent The Eddington limit for a spherical, non-rotating, homogenous object was discussed from a modern perspective by Joss, Salpeter \& Ostriker \cite{JosSalOst74}. In addition to the discussion given there, let us note that, obviously, the limit for radiative power may increase above the standard Eddington limit, in objects with a gravitational cross section greater than the standard one, or with a radiative cross section smaller than the standard one\footnote{In no circumstances $L_{\rm Edd}$ can grow above $L_{\rm Planck}$. I noticed \cite{Abr......88} that the two limits are equal when $M \approx N_0 m_{\rm P}$, with $N_0 = N_{\rm Dirac}^2/3$, where the Dirac number $N_{\rm Dirac} = e^2/G m_{\rm e} m_{\rm P}$ equals the ratio of Coulomb's to Newton's force between electron and proton. Then $N_0 \approx N_{\rm Edd} \equiv 136 \times 2^{256} \approx 1.6 \times 10^{79}$, where $N_{\rm Edd}$ is the Eddington number that played an important role in Eddington's {\it Fundamental Theory} \cite{Edd......28}. The number was introduced by Eddington's immortal statement, {\it I believe there are 15 747 724 136 275 002 577 605 653 961 181 555 468 044 717 914 527 116 709 366 231 425 076 185 631 031 296 protons in the universe and the same number of electrons}. One thus may write, $L_{\rm Edd} = (N/N_{\rm Edd})\,L_{\rm Planck}$. A very Eddingtonish connection indeed, embracing his luminosity and his number, of which he was not aware.},

\begin{equation}
\label{poss}
\Sigma_{\rm grav}^* > \Sigma_{\rm grav} \equiv 4 \pi R_G^2, 
~~\Sigma_{\rm rad}^* < \Sigma_{\rm rad} \equiv {M \over m_{\rm P}}\,\sigma_T\, .
\end{equation}

\noindent Several types of astrophysical sources give a clear observational evidence for super-Eddington luminosities. I will mention here only two specific types that provide examples for the two possibilities listed in (\ref{poss}). 

{\it Novae outbursts.} They are due not to accretion but to thermonuclear power \cite{Sha......98}. Luminosity increases within a few hours by factors of $10^4$, and stay for a very long duration at a clearly super-Eddington level, $L > 10\,L_{\rm Edd}$. Shaviv \cite{Sha......98}, \cite{Sha......01} shows that the observed increase of the radiative power over the Eddington limit, may be attributed to $\Sigma_{\rm rad}^* < \Sigma_{\rm rad}$, because in a locally inhomogeneous medium, the ratio of averaged radiation force to emitted flux, $F_{\rm rad}/f_{\rm rad} \equiv \sigma_{\rm rad}^*/c < \sigma_T/c$. 

{\it X-ray binaries.} Some of them show super-Eddington luminosities, powered, most likely, by super-Eddington accretion: in high-mass binaries undergoing a thermal-time-scale mass transfer, e.g. in SS433, \cite{KinTaaBeg20}, and in the low-mass binaries during long-lasting transient outbursts, e.g. in GRS 1915+105, \cite{MirChaRod98}. 

The rest of my review is devoted to the discussion of a possibility that the reason for super-Eddington luminosities in the X-ray binaries and similar objects may be the fast differential rotation of accretion disks present in these sources. It strengthens the effective gravity, so that $\Sigma_{\rm grav}^* > 4 \pi R_G^2$. 

\section{Eddington limit for rotating bodies}

I now consider the Eddington limit for a perfect-fluid rotating body in equilibrium, following the line of arguments first presented by Abramowicz, Calvani \& Nobili \cite{AbrCalNob80}. Having in mind most general astrophysical applications, I will consider two topologically different cases: 

{\it Rotating stars.} The surface of the star has a topology of the sphere. The whole mass $M$ is included in the sphere. 

{\it Accretion disks.} The surface of the disk has a topology of the torus. The mass of the disk $M_{\rm disk}$ is contained in the torus, but the mass $M_{\rm centr}$ of the central black hole is outside. The total mass $M = M_{\rm centr} + M_{\rm disk} \approx M_{\rm centr}$, because $M_{\rm centr} \gg M_{\rm disk}$. In accretion disk theory it is customary to neglect the mass of the disk, so formally $M = M_{\rm centr}$, and $M_{\rm disk} =0$. The Eddington limit always refers to the total mass $M$.

Let ${\bf f}_{\rm rad}$ be the local flux of radiation somewhere at the surface of the body, $S$. The corresponding radiative force is ${\bf F}_{\rm rad} = (\sigma_{\rm rad}/c){\bf f}_{\rm rad}$. Let ${\bf F}_{\rm eff} = {\bf F}_{\rm grav} + {\bf F}_{\rm rot}$, be the effective gravity force, with ${\bf F}_{\rm grav} = m \bnabla \Phi$ being the gravitational force ($\Phi$ is the gravitational potential), and with ${\bf F}_{\rm rot} = m (\Omega^2 r)\,{\bf e}_{\rm r}$ being the centrifugal force ($\Omega$ is the angular velocity, $r$ is the distance from the axis of rotation, and ${\bf e}_{\rm r} = \bnabla r$ a unit vector showing the off-axis direction).

The necessary condition for equilibrium is ${\bf F}_{\rm rad} < {\bf F}_{\rm eff}$. From this one deduces the Eddington limit for rotating perfect-fluid bodies,

\begin{equation}
L = \int_{\rm S} {\bf f}_{\rm rad}\cdot d{\bf S} = {c \over \sigma_{\rm rad}}\int_{\rm S} {\bf F}_{\rm rad}\cdot d{\bf S} \,<\, {c \over \sigma_{\rm rad}}\int_{\rm S} {\bf F}_{\rm eff}\cdot d{\bf S} \equiv L_{\rm Edd}^{\rm rot}
\end{equation}

\noindent Using the Gauss theorem to transform the surface integral of ${\bf F}_{\rm eff}$ into a volume integral of $\bnabla \cdot {\bf F}_{\rm eff}$, Poisson equation $\bnabla^2 \Phi = 4\pi G\rho\,$, and introducing the specific angular momentum $\ell = \Omega r^2$, one gets after twenty or so lines of simple algebra,

\begin{equation}
L_{\rm Edd}^{\rm rot} 
= L_{\rm Edd} \left [\,X^2_{\rm mass} + X^2_{\rm shear} - X^2_{\rm vorticity}\,\right ],
\end{equation}

\noindent where the dimensionless number $X^2_{\rm mass}$ depends on whether the body is a star, or an accretion disk,

\begin{equation}
X^2_{\rm mass} = {1 \over M } \int_{\rm V} \rho dV =
\left \{
\begin{array}{l}
1 \quad {\rm for~stars,}
\\
\\
0 \quad {\rm for~accretion~disks,}
\end{array}
\right .
\end{equation}

\noindent and where $X^2_{\rm shear}$, $X^2_{\rm vorticity}$ are dimensionless, necessarily positive quantities, representing shear and vorticity integrated over the whole volume of the body, 

\begin{equation}
X^2_{\rm shear} = {1\over {16\pi G M}}\int_{\rm V} r^2 ({\bnabla} \Omega \cdot \bnabla \Omega)\,dV,
\end{equation}
\begin{equation}
X^2_{\rm vorticity} = {1\over {16\pi G M}}\int_{\rm V} r^{-2} (\bnabla \ell \cdot \bnabla \ell)\,dV.
\end{equation}

\noindent Shear increases the Eddington limit, and vorticity decreases it. 

The rotation of astrophysical objects is far from simple, but an insight 
could be gained by considering a simple power law for the angular momentum distribution, $\ell(r, z) = \ell_0 r^a$, with $\ell_0$ and $a$ constant. Rigid rotation has $a=2$, Keplerian rotation $a=1/2$, and constant angular momentum rotation $a=0$. It is $X^2_{\rm shear}/X^2_{\rm vorticity} =(a-3)^2/a^2$. This means that $X^2_{\rm shear} > X^2_{\rm vorticity}$ when $a < 3/2$.  

Rotating stars have $X^2_{\rm shear} < X^2_{\rm vorticity}$, because they rotate almost rigidly. Thus, for rotating, radiation pressure supported stars, $L \approx L_{\rm Edd}^{\rm rot} < L_{\rm Edd}$ {\it always}. Contrary to this, constant angular momentum tori are dominated by shear, $X^2_{\rm shear} \gg 1 \gg X^2_{\rm vorticity}$ and consequently, when they are radiation pressure supported, $L \approx L_{\rm Edd}^{\rm rot} \gg L_{\rm Edd}$. 

\section{Polish doughnuts}

For a simplicity of presentation, let me assume that $\ell = \ell_0 = {\rm const}$, i.e. that the angular momentum is constant in the whole body.
One does not know a priori what is the actual distribution of the angular momentum inside the body, as this depends on the nature of viscosity. Paczy{\'n}ski explained why a physically realistic distribution of angular momentum must indeed be close to $\ell = {\rm const}$ in the inner part of the flow, and why most likely it approaches the Keplerian distribution, $\ell(r,z) = \ell_K(r) \equiv (GMr)^{1/2}$ far away from the center. 

Adopting Paczy{\'n}ski's argument, one may assume that the local physical properties of the innermost part of black hole accretion flows are rather well described by the model with $\ell = {\rm const}$, but one must be careful with the physical interpretation of the $\ell = {\rm const}$ assumption at large radii. I will return to this point later in this section.

Assuming constant angular momentum, and using Newton's expression for gravitational potential $\Phi = -GM/( r^2 + z^2 )^{1/2}$ in cylindrical coordinates $[\,r, z, \phi$\,], I deduce the shape of equipressure surfaces $P = {\rm const}$ from the Bernoulli equation, 

\begin{equation}
\label{equipotential-1}
-{GM \over {\left( r^2 + z^2 \right)^{1\over 2}}} + {1\over 2} {\ell^2 \over r^2} + W(P) = {\rm const}.
\end{equation}

\noindent Here $W(P)= -\int dP/\rho$. This equation cannot be obeyed at the rotation axis (where $z \not = 0$, $r = 0$), which has an obvious, but important, consequence: no equipressure surface cross the axis. For a constant angular momentum fluid, equipressure surfaces must be toroidal, or open. The marginally open surface has just one point $(r=\infty, z=0)$ at infinity. This particular surface encloses the {\it largest possible} torus. From (\ref{equipotential-1}) it is obvious, that in this case $W(P) = W(0) = 0$. 

Maximal pressure locates at a circle $z=0, r=r_0 = GM/\ell_0^2$. Using the radius $r_0$ as a scale, $\xi = r/r_0$, $\eta = z/r_0$, $w = W/(GM/r_0)$, one may write equation (\ref{equipotential-1}) in the dimensionless form, and solve for $\eta = \eta (\xi; w)$ to obtain the explicit form of all equipressure surfaces, 

\begin{equation}
\label{equipotential-2}
\eta^2 = Q^2(\xi) \equiv 4\xi^4 \left( 1 - 2 \xi^2 w \right)^{-2} - \xi^2,~~-1/2 \le w \le w_S.\,
\end{equation}

\noindent The value $w = -1/2\,$ gives the location of the center, and $w = w_S \le 0\,$ the location of the surface. For the {\it slender torus} $w_S \approx -1/2$, and the {\it fat torus} $w_S \approx 0$.

\begin{figure}[ht]
\begin{center}
\includegraphics[width=.48\textwidth]{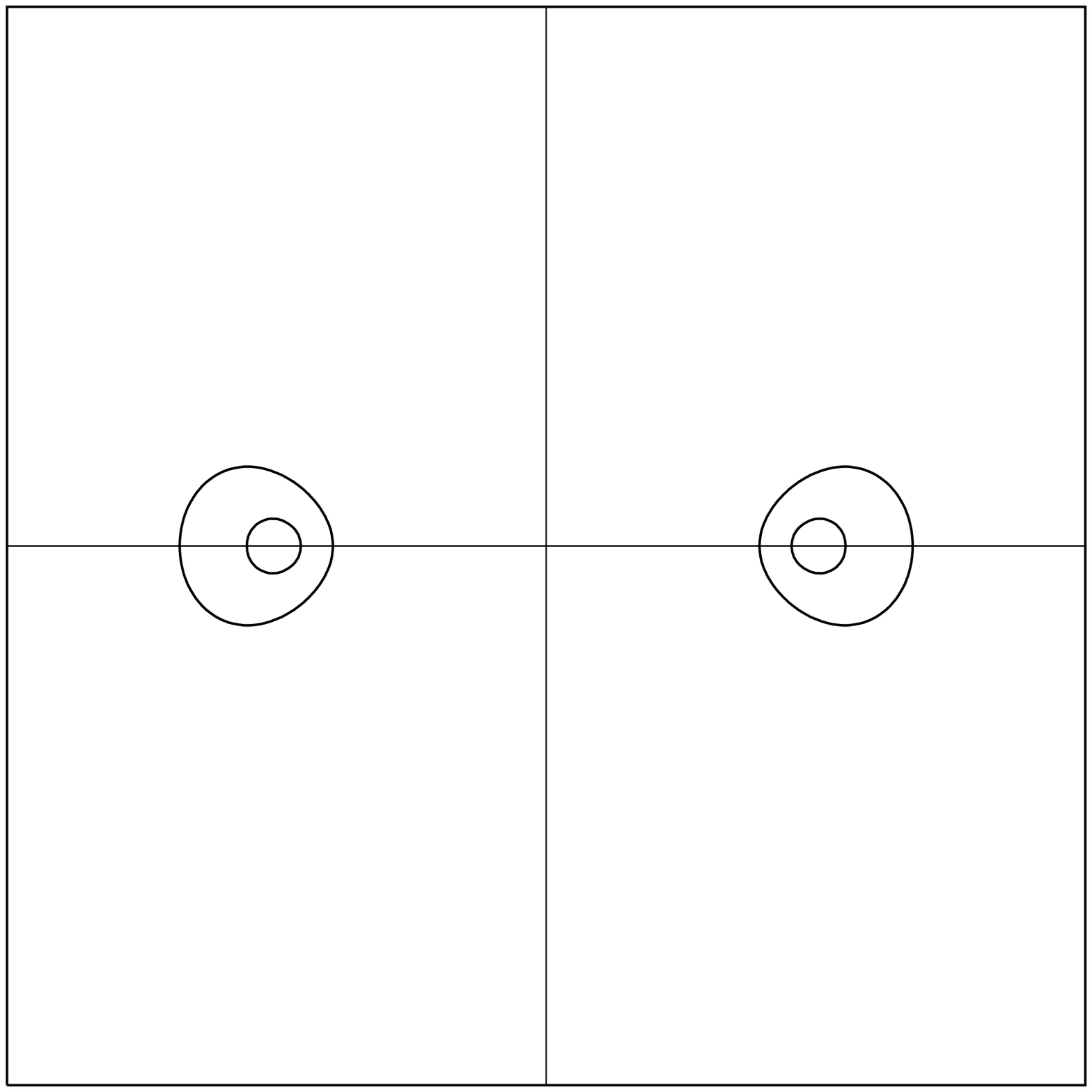}
\hfill
\includegraphics[width=.48\textwidth]{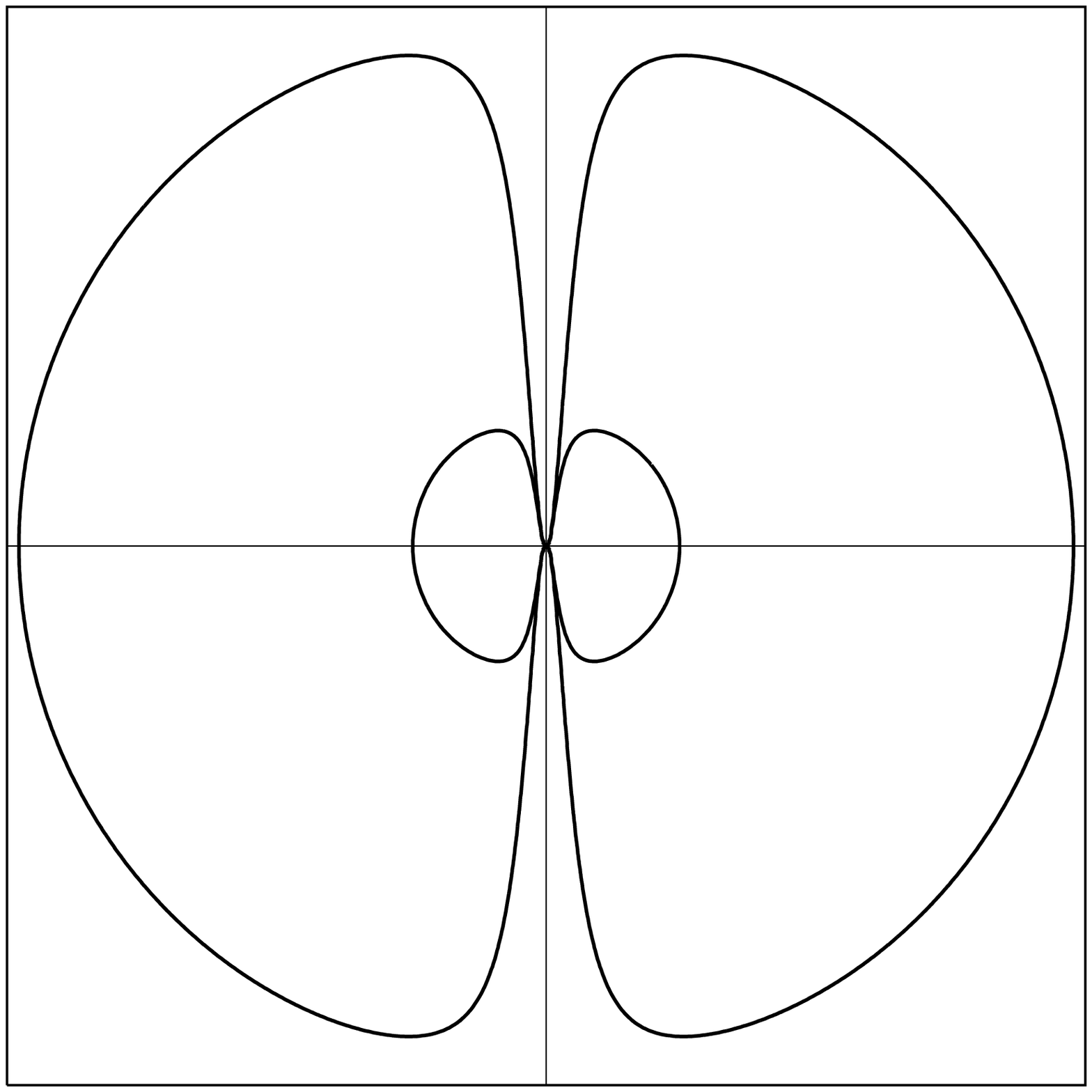}
\end{center}
\caption[]{Contours of equipotential surfaces on the meridional cross section of a constant angular momentum torus. {\it Left}: a slender torus, $w_S \approx -1/2$. The contours approach circles around the locus of the maximum pressure, at $\xi =1$. {\it Right}: a fat torus $w_s \approx 0$. The contours tend to concentric circles with the center at $\xi =0$. Close to axis they change into a pair of conical funnels.}
\label{slender-fat}
\end{figure}

\subsection{The slender torus}

One may introduce toroidal coordinates centered at the circle of the maximal pressure by a coordinate transformation, $x = [\eta^2 + \xi_*^2]^{1/2}$,
$\alpha = \tan^{-1} (\eta/\xi_*)$. Here $\xi_* \equiv 1 - \xi$. It is then not difficult to show that in the slender torus limit, i.e. when $w + 1/2 = \epsilon \ll 1$, it is $w = w(x) + {\cal O}^2 (\epsilon)$. This means that at the meridional section of the torus, equipressure surfaces are concentric circles, as first noticed by Paczy{\'n}ski \cite{MadPac...77}. This additional symmetry was used by Blaes \cite{Bla......85} to obtain a general analytic solution for all possible slender torus oscillation modes. In particular, Blaes demonstrated that there exists a set of non-axisymmetric (i.e. with $m \not = 0$) global modes $\delta X_{mn} = F_{mn}(x, \alpha)\exp (-i \omega_{mn} t + m \phi)$ of the slender torus oscillations, with  the eigenfrequencies,

\begin{equation}
\label{omer-unstable}
\omega_{mn} = - m \Omega_K \left [ 1 + i \epsilon \left ( { 3 \over {2n + 2}} 
\right )^{1\over 2} \right] + {\cal O}^2 (\epsilon), ~~n=1,2,3,...
\end{equation}

\noindent Because ${\Im}(\omega_{mn}) < 0$ these oscillations are
unstable. The growth rate of the instability is $\sim m\Omega_K$. Here
$\Omega_K$ is the Keplerian frequency at the torus center, and thus
the instability is a {\it dynamical} one. Indeed, this is the famous
instability, discovered in the seminal papers by Papaloizou and
Pringle \cite{PapPri..84a}, \cite{PapPri..84b}. 

\subsection{The fat torus}

It is easy to understand shapes of equipotential surfaces in a fat torus (shown in Figure \ref{slender-fat}). Very far from the axis of rotation one has $|2w\xi | \gg 1$. Inserting this into (\ref{equipotential-2}) one gets $\eta^2 + \xi^2 = 1/w^2$ which means that far from the rotation axis the equipotential surfaces are spheres with radius $1/|w|$, and that the outer radius of the torus is at $r_{\rm out} = r_0/|w_S|$. Spherical equipotentials are in accord with the fact that very far from the axis, $\xi \gg 1$, the centrifugal force $\sim \ell_0/\xi^3$ is negligible with respect to the gravitational force $\sim GM/\xi^2$. Therefore, the effective gravity is determined by Newton's attraction alone, as for spherical stars. For the same reason, the radiation power from this part of the surface of a fat, radiation pressure supported torus (i.e. Polish doughnut) is {\it one Eddington luminosity}, the same as from a spherical non-rotating, radiation pressure supported star\footnote{M.~Rees \cite{Ree......80} told me in 1980 that this implied that Polish doughnuts, with at least one Eddington luminosity from whatever direction, were obviously too luminous to describe very ``dim'' active galactic nuclei, such as radio galaxies: ``while apparently supplying tremendous power to their extended radio-emitting regions, the nuclei of most radio galaxies emit little detectable radiation.'' He and his collaborators at Cambridge found later a possible solution to this puzzle in terms of the {\it ion pressure supported tori} \cite{BlaBegPhi82}. The ion tori have the same shapes as the Polish doughnuts, in particular funnels, but much lower, indeed very sub-Eddington, luminosities. The power in jets comes from tapping rotational energy of the central black hole by the Blandford-Znajek mechanism \cite{BlaZna...77}.}.

Note, however, that the asymptotically spherical shape of a fat torus is a direct consequence of the assumption $\ell(r,z) ={\rm const}$, which was made ad hoc. If one adopts a more physically realistic assumption that asymptotically $\ell(r,z)=\ell_K$, one may use the standard Shakura-Sunyaev model in its radiation pressure version, to get the asymptotic shape of the fat torus,

\begin{equation}
\label{asymptotic-thickness}
z_{\infty} = {{3\sigma_T} \over {8\pi c m_p}}\, {\dot M}.
\end{equation}

Closer to the axis, $|2w\xi | \sim 1$, which means that $\xi^2 \sim -1/2w$ and this together with (\ref{equipotential-2}) gives $\eta^2/\xi^2 = (1/|w|) - 1 = (1/\sin^2 \theta) - 1$, i.e. that closer to the axis, equipotential surfaces corresponding to $w \sim w_S \sim 0$ have conical shapes with the half opening $\theta \sim \sqrt w$. The surfaces are highly non spherical because centrifugal force dominates. Integrating effective gravity along the conical funnel is elementary, and one gets \cite{JarAbrPac80} that $L^{\rm rot}_{\rm Edd}/L_{\rm Edd} = (1/2) \ln (1/|w_S|)$. This estimate may be used to find the total luminosity for the radiation pressure supported fat torus. i.e. a Polish doughnut, 

\begin{equation}
\label{logarithm}
{L_{\rm total} \over L_{\rm Edd}} \approx {L_{\rm Edd}^{\rm rot} \over L_{\rm Edd}} \approx {1\over 2} \ln \left({r_{\rm out}\over r_{0}}\right ) = 1.15\,\log\left ({r_{\rm out} \over r_0}\right ).
\end{equation}

\noindent It should be clear from our derivation, that the logarithmic 
scaling of the luminosity with the torus size is a genuine property of 
the Polish doughnuts, including those that have a non-constant angular momentum distribution. The logarithm in (\ref{logarithm}) is of a crucial importance, as it prevents astrophysically realistic doughnuts (i.e. with $r_{\rm out}/r_{0} < 10^6$, say) to have highly super-Eddington luminosities. Thus, the theory predicts that for such ``realistic'' fat tori, only a slightly super-Eddington total (isotropic) luminosities, $L_{\rm total} \ge 7\,L_{\rm Edd}$, may be expected. 

However, because the funnels have solid angles  $\theta^2 \sim r_{0}/r_{\rm out }$, radiation in the funnels may be, in principle, collimated to highly super-Eddington values $L_{\rm coll} /L_{\rm Edd} = \Theta \sim r_{\rm out}/r_{0} \gg 1$. This simple estimate agrees with a more detailed modelling of the Polish doughnuts radiation field by Sikora \cite{Sik......81} and Madau \cite{Mad......88} who obtained $\Theta \le 10^2$ for disks with $r_{\rm out}/r_{0} \sim 10^2$. A typical value that follows from observational estimates for non-blazar active galactic nuclei, e.g. by Czerny and Elvis \cite{CzeElv...87} and Malkan \cite{Mal......89}, is $\Theta \sim 10$, but of course for blazars and other similar sources (e.g. for ULXs, if they are powered by stellar mass black holes, as argued by King \cite{Kin......02}), it must be $\Theta \gg 10$.

Such high values of $\Theta$ are also consistent with the idea, suggested by Pa\-czy{\'n}\-ski \cite{Pac......80} and independently by Lynden-Bell \cite{Bel......82}, that relativistic electron-positron $e^-e^+$  jets may be very effectively accelerated by the radiation pressure in the fat tori funnels\footnote{Lynden-Bell called this an ``entropy fountain''.}. Note, that if the flux in the funnel is $\Theta$ times the Eddington flux, the $e^-e^+$ plasma feels the ``effective'' radiative force corresponding to the Eddington ratio $m_p/m_e \approx 10^3$ times greater\footnote{Note that in Shaviv's explanation of the situation during nova outburst, the luminosity is {\it physically} very close to the Eddington one (in the sense of Sikora's definition), although its value is greater than the standard Eddington limit. For the $e^-e^+$ jets the situation is very much different (opposite):  the luminosity is {\it physically} very super-Eddington, although it may be just slightly above the standard limit.}. Detailed calculations (e.g. \cite{AbrPir...80}, \cite{Sik......81}, \cite{SikWil...81}) demonstrated that indeed the $e^-e^+$ jets may be accelerated in funnels up to the Lorentz factor $\gamma \le 5$. However, if jets are initially pre-accelerated by some black-hole electrodynamical processes (such as the Blandford-Znajek mechanism \cite{BlaZna...77}) to highly relativistic velocities $\gamma > 10^6$, they will be decelerated in the funnels by the Compton drag, reaching the asymptotic Lorentz factor \cite{AbrEllLan90},

\begin{equation}
\label{lanza}
\gamma_{\infty} = \left ( \Theta \,{m_p \over m_e} \right )^{1\over 3} =
10 \,\Theta^{1\over 3}.
\end{equation}

\noindent Observations tell that $\gamma_{\infty} < 10^2$, and thus equation (\ref{lanza}) suggests that $\Theta < 10^3$.

\subsection{Rise and Fall of the Polish Doughnuts}

In the time of their d{\'e}but, Polish doughnuts could theoretically confirm the observed super-Eddington luminosities, highly collimated beams of radiation, and perhaps even the relativistic speeds of jets, fulfilling another principle attributed to Eddington \cite{Wei......92}: {\it one should never believe any experiment until it has been confirmed by theory.}  These virtues attracted initially some interest of the astrophysical community, but the interest has quickly drained with the discovery of the Papaloizou-Pringle instability. It was thought that Polish doughnuts must necessarily suffer from the instability and thus, in reality, they cannot exist. The important discovery by Blaes \cite{Bla......87} that the Roche lobe overflow stabilizes Polish doughnuts against the Papaloizou-Pringle instability, came too late --- in the advent of numerical simulations of black hole accretion flows. Too late, because numerical simulations rediscovered and absorbed many of the Polish doughnuts results. Today, these results exist in the consciousness of many astrophysicists as a set of several numerically established, important but unrelated facts. They do not form a consistent scheme that the Polish doughnuts once offered: clear, simple, following directly from the black hole physics\footnote{There are at least three brilliant and very pedagogical expositions of the Polish doughnuts scheme: two by Paczy{\'n}ski himself, \cite{Pac......82}, \cite{Pac......98}, and one in the text book by Frank, King and Raine \cite{KinFraRai02}.}.  

In the rest of my review, I recall the most fundamental of these results, almost totally forgotten today: the relativistic Roche lobe overflow mechanism. The mechanism not only stabilizes Polish doughnuts against the Papaloizou-Pringle and other instabilities. It also assures that highly super-Eddington accretion rates always imply a very low efficiency of accretion\footnote{So{\l}tan's well known argument \cite{Sol......82} shows that {\it on average} the AGN efficiency cannot be small. The argument  does not exclude a possibility of a flip-flop sequence of periods of low and high efficiency, rather like in \cite{Li.Pac...00} for thin disks. Although the possibility of a flip-flop behavior with timescale $\sim 10^4$ yrs (for AGN) has some observational back-up \cite{Abr.....a85}, it was never studied sufficiently deeply.}.  

\section{Efficiency of black hole accretion}

I like to illustrate the physical meaning of accretion disk efficiency in terms of the Bekenstein engine\footnote{Here I borrow a short description of the engine given in \cite{NarIguAbr03}.}. Bekenstein \cite{Bek......72} discussed a black hole engine that converts mass to energy with a perfect efficiency, following an earlier unpublished remark by Geroch \cite{Ger......71}. The engine works by slowly lowering on a strong wire a mass $m$ into a black hole with the Schwarzschild radius $R_S = 2\,R_G = 2GM/c^2\,$ and mass $M$.

As the mass is lowered to a radius $R$, the energy $E(R)$ measured at infinity is gravitationally redshifted and thus goes down relative to the initial energy $E_0 = mc^2$. The change in energy equals to the mechanical work done by the wire back at the engine. The efficiency $e \equiv [{E_0 - E(R)]/E_0}$ may be calculated from the gravitational redshift formula, $e = 1 - ( 1 - R_S /R )^{1/2}$. If the mass could be lowered to the horizon $R \rightarrow R_S$, the efficiency would go to $1$. Of course, the mass cannot touch the horizon, because the tension in the wire would be then infinite, and even the strongest wire would break. Real wires, that may sustain only a finite tension (see e.g. Gibbons \cite{Gib......72}), would break a finite distance $R=R_{\it in}$ from the horizon. In a region below the breaking radius, $R<R_{\rm in}$, the mass falls down freely, keeping its energy measured at infinity unchanged. Therefore, the efficiency is determined by the radius $R=R_{\rm in}$ at which the wire breaks, $e \equiv [{E_0 - E(R_{\rm in})]/E_0}$.

\begin{figure}[ht]
\begin{center}
\includegraphics[width=.46\textwidth]{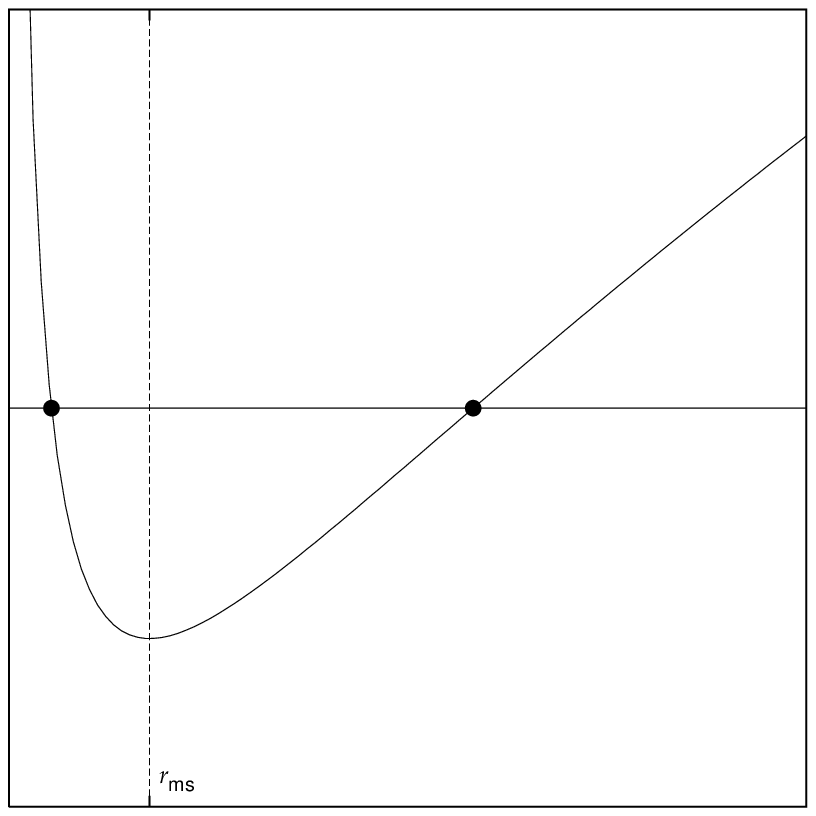}
\hfill
\includegraphics[width=.46\textwidth]{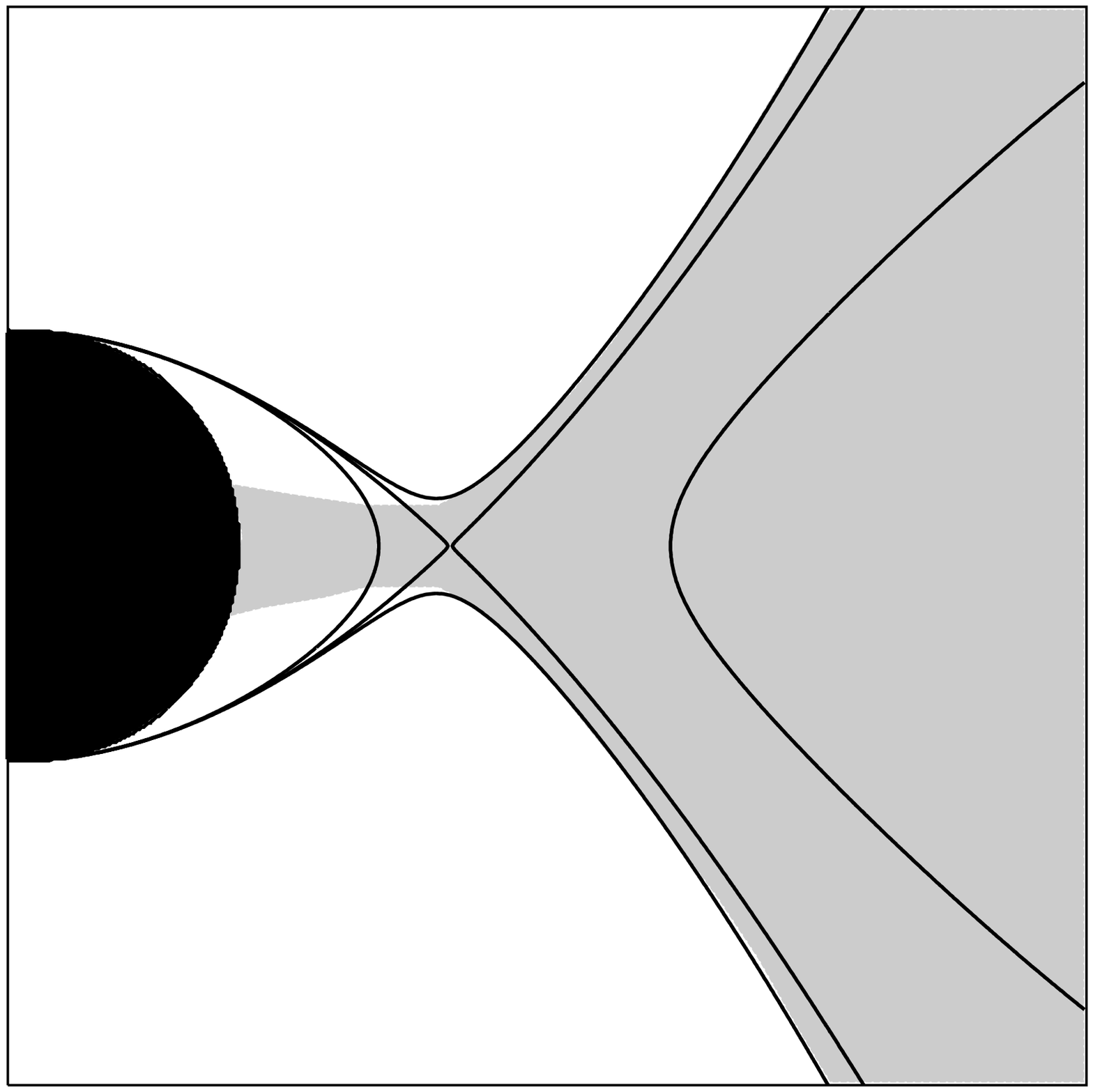}
\end{center}
\caption[]{Left Figure shows that angular momentum in the fat torus crosses twice the Keplerian angular momentum. Right Figure, shows schematically that if the the doughnut slightly overflows the Roche lobe, a dynamical mass loss will occur.}
\label{Roche}
\end{figure}

With centrifugal support playing a role of the rope, accretion's efficiency is also determined by the radius $R=R_{\rm in}$ at which the centrifugal support breaks. 

Consider a collection of free particles on Keplerian circular orbits around a black hole. Small viscosity will slowly remove angular momentum, so particles will slowly drift inward. However, in the region $R < R_{\rm ms}$, there are no stable circular orbits, and therefore the centrifugal support breaks: the particles free-fall into the black hole. Thus, similarly as in the case of the Bekenstein engine, the Keplerian binding energy of particles at the location of the marginally stable orbit, $R_{\rm ms}$ (also called ISCO), determines the efficiency of accretion.

\subsection{The relativistic Roche lobe overflow}

The location of $R_{\rm ms}$ is determined by the minimum of the Keplerian angular momentum $\ell = \ell_K$, shown in Figure \ref{Roche}, left. For a constant angular momentum fluid torus, $\ell = \ell_0$, the combined pressure and centrifugal support breaks at the radius $R_{\rm in} < R_{\rm ms}$, defined by $\ell_K = \ell_0$ and shown in the same Figure, by the dot left of $R_{\rm ms}$ on the angular momentum distribution. 

At this location, as discovered by Abramowicz, Jaroszy{\'n}ski and Sikora \cite{AbrJarSik80}, one of the equipotential surfaces, called the Roche lobe, self-crosses as shown in Figure \ref{Roche} (right). It should be obvious that if the fluid distribution overflows the Roche lobe, a dynamical mass loss must occur: at the the circle $R = R_{\rm in}$ the centrifugal and pressure support breaks down, and fluid if free falling in the region $R < R_{\rm in}$. 

Again, the efficiency is determined by the Keplerian binding energy at breaking point, $R_{\rm in}$, called the {\it inner edge} of the accretion disk. This name is the source of a confusion, as some wrongly imagine that the name implies that at the inner ``edge'' velocity, density, and pressure must experience a jump, or a sudden change. Of course they do not. What {\it does} change in the region close to $R_{\rm in}$ is the nature of the flow: from sub-sonic to super-sonic. Note also that for low angular momentum flows, with $\ell < \ell_K$ everywhere, there is no Roche lobe, and therefore the inner edge cannot be sensibly defined. Such flows differ considerably from the Polish doughnuts, see e.g. \cite{BegMei...82}. Typically, they assume high viscosity, while Polish doughnuts assume low viscosity, less than about $0.03$ in terms of the Shakura-Sunyaev $\alpha$-parameter. 

\begin{figure}[ht]
\begin{center}
\includegraphics[width=.47\textwidth]{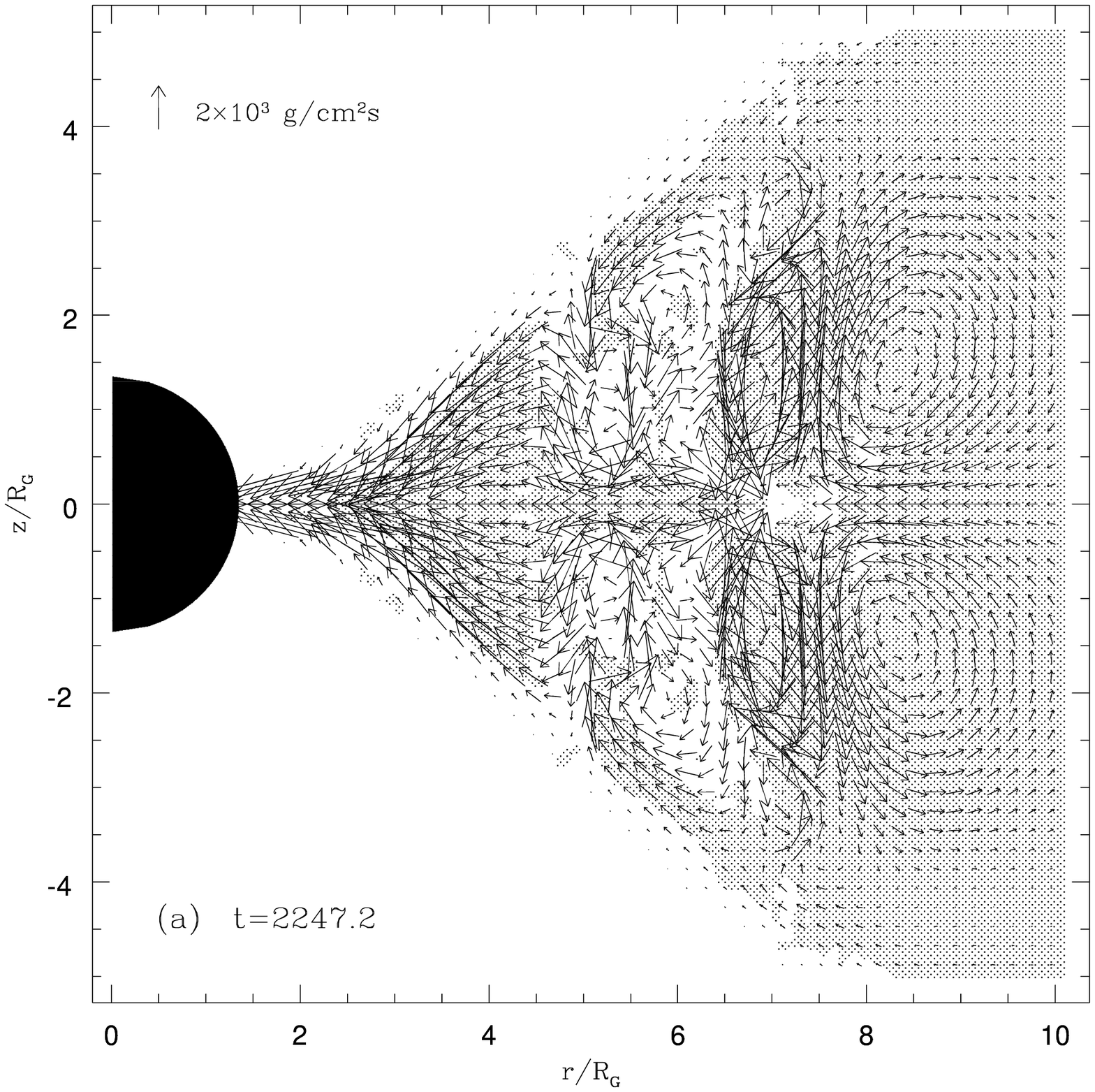}
\hfill
\includegraphics[width=.48\textwidth]{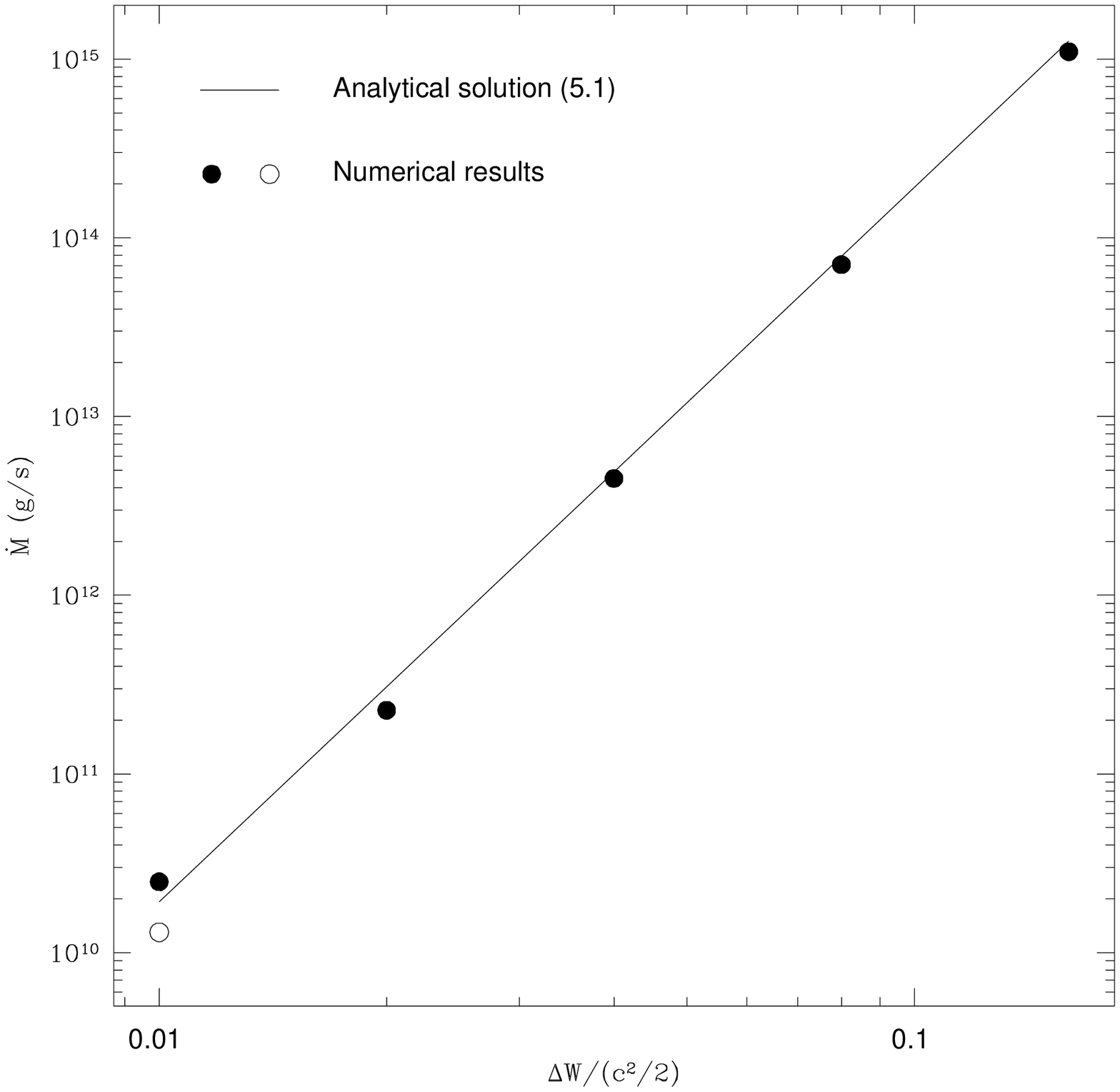}
\end{center}
\caption[]{Left Figure shows a snap-shot of the velocity field inside a time-dependent accretion black hole flow. The nature of the flow clearly changes at the inner edge (although physical characteristics do not experience a jump there). Far away from the inner edge, the flow pattern is very complicated, showing back flows, vortexes, and circulations in convective cells. Close to the inner edge, the Roche lobe overflow mechanism organizes the flow into a highly ordered, almost radial pattern. Right Figure shows that despite all complications, the Roche lobe overflow accretion rate self-regulates according to the simple analytic formula (\ref{Euler-Gamma}). The points are calculated in supercomputer simulations \cite{IguCheAbr96} of 3D, viscous, time dependent, black hole accretion flows with different parameters and different boundary conditions. The straight line is given by the formula (\ref{Euler-Gamma}). One should appreciate the nearly perfect agreement between this simple analytic formula and the most sophisticated, indeed today's state-of-art, numerical simulations. Both figures are taken from \cite{IguCheAbr96}.}
\label{roche-igor}
\end{figure}

It was shown \cite{KozAbrJar80} that the inner edge must locate between the 
marginally stable, and marginally bound orbits,
\begin{equation}
\label{inner-edge-location}
R_{\rm mb} < R_{\rm in} < R_{\rm ms}.
\end{equation}

\noindent The marginally bound orbit at $R_{\rm mb}$ has the same binding energy at infinity, $W(\infty) = W(R_{\rm mb})$. This has two important consequences: the efficiency of a torus with $R_{\rm in} = R_{\rm mb}$ is zero, {\it and} such a torus must have its outer radius at a very large distance, formally at infinity, $R_{\rm out} = \infty$. We have previously shown that the total luminosity of a fat torus increases logarithmically with the outer radius. This, combined with the accretion rate being proportional to the size (\ref{asymptotic-thickness}), yields the logarithmic behavior $L_{\rm total}/L_{\rm Edd} \sim 
\ln ({\dot M}/{\dot M}_{\rm Edd})$. Indeed, Paczy{\'n}ski \cite{Pac.....b80} found that all fat disks calculated by him obey

\begin{equation}
\label{paczynski-logarytm}
{L_{\rm total} \over L_{\rm Edd}} = 4.8 \log \left ( {{16\,\dot M} \over {\dot M}_{\rm Edd}}\right ).
\end{equation}

\noindent This shows that the super-critical accretion is qualitatively different from the standard, sub-Eddington, thin disk accretion. For the standard thin disks, the total luminosity is directly proportional to the accretion rate $L_{\rm total}/L_{\rm Edd} = e(r_{\rm ms}){\dot M}/{\dot M}_{\rm Edd}~,e(r_{\rm ms})={\rm const}$. 

The Roche lobe overflow mechanism self-regulates the accretion rate, which for a polytropic equation of state, $P = K\,\rho^{1 + 1/n}$ equals, 

\begin{equation}
\label{Euler-Gamma}
{\dot M} = A(n)K^{-n} {R_{in} \over {\Omega_K (R_{in})}}{|\Delta W |}^{n+1},~~
\Delta W \equiv W_S - W_{in}
\end{equation}

\noindent with an analytic expression for $A(n)$ explicitly known in terms of the Euler gamma function, $\Gamma(n)$. Because $n=3$ for a radiation pressure supported gas, the self-regulation imposed by (\ref{Euler-Gamma}) is a very strong one, ${\dot M} \sim |\Delta W |^4$. Suppose, that in the region close to the inner edge, there is a fluctuation in the thermal balance, causing overheating. This will induce an expansion, and an increase in the Roche-lobe overflow, which in turn will increase the accretion rate, i.e. the mass loss from the region. The heat contained in the mass that is lost will cool down the region, thus assuring thermal stability. I found \cite{Abr......81} that this mechanism of advective cooling caused by the Roche lobe overflow, always stabilizes the innermost region of {\it any} accretion disk (thin, slim, thick, adaf) against thermal and viscous instability. Roughly speaking, the {\it local} instabilities have no chance to grow in the innermost region of accretion disks, because they are quickly washed away by advection caused by the Roche lobe overflow.

\noindent Stabilization of the {\it global} Papaloizou-Pringle instability by the Roche lobe overflow, found by Blaes \cite{Bla......85}, has a different physical reason: almost a perfect reflection of the Papaloizou-Pringle mode at the inner disk edge is a necessary ingredient of the instability. The reflection is not possible in the fast (transonic) flow induced by the Roche lobe overflow. Thus, the non-realistic, non-accreting Polish doughnuts indeed always suffer from the Paploizou-Pingle instability. In a more realistic model of a Polish doughnut, that includes the Roche lobe overflow, the Paploizou-Pringle instability does not operate.

\section{Discussion}

\subsection{Paczy{\'n}ski's pseudo-potential}

An accurate and elegant model for a non-rotating black hole's gravity was introduced by Pa\-czy{\'n}\-ski \cite{PacWii...80} it terms of a Newtonian gravitational ``pseudo-potential'', 

\begin{equation}
\label{paczynski-wiita}
\Phi_{PW} (R) = - { GM \over {R - R_S}},~R_S = {2GM \over c^2}.
\end{equation}

\noindent The Paczy{\'n}ski model is remarkably successful: numerous authors used it in their calculations of black hole accretion flows. This clever idea cannot be, however, applied for: (a) rotating black holes, because of the Lense-Thirring effect\footnote{Indeed, all suggested ``Kerr pseudo-potentials'', are neither elegant, nor practical.}, and (b) self-gravitating fluids, as $\nabla^2 \Phi_{PW} \not = 0$.

\subsection{Standard thin disks are inconsistent with ${\dot M} > {\dot M_{Edd}}$}

Properties of accretion flows with ${\dot M}>{\dot M}_{\rm Edd}$ are sometime discussed in terms of the standard thin Shakura-Sunyaev model. However, the standard thin disk is inconsistent with the super-Eddington accretion. Indeed, the radiative force cannot be greater than the effective gravity force, and at the surface of the standard disk this condition yields,  

\begin{equation}
\label{Pacz1}
F_{\rm grav} = {{GMm_{\rm P}} \over {r^2}}\left({H \over r}\right) > F_{\rm rad} = {\sigma_T \over c} f_{\rm rad} =  
{\sigma_T \over c}{{3GM{\dot M}} \over {8\,\pi r^3}} \left[ 1 - \left ( {{3r_G} \over r} \right ) \right ]^{1 \over 2}\, . 
\end{equation}

\noindent From (\ref{Pacz1}) Paczy{\'n}ski \cite{Pac......80} derived $(H/r)_{max} > {\dot M} /{\dot M}_{\rm Edd}\,$. This means that for a super-Eddington accretion  with ${{\dot M}/{\dot M}_{\rm Edd}} > 1\,$ it also must be $H/r >1$. However, the standard model is a 1D approximation to the 3D accretion physics and assumes $H/r \ll 1$. Its structure equations contain only zero and first order terms in $H/r$. Therefore, the super-Eddington accretion is outside of the standard model applicability range. Contrary to this, slim disks and Polish doughnuts are suitable to describe super-Eddington accretion: slim disks are accurate up to the second order terms in $H/r$, and Polish doughnuts are described by the full 3D equations (i.e. contain $H/r$ terms of {\it all} orders).

\subsection{Adios: reasonable conclusions from false arguments}

The Blandford \& Begelman ``adios'' paper \cite{BlaBeg...99} claims that accretion with small efficiency (Polish doughnuts, adafs) must {\it necessarily} experience strong outflows, {\it because} matter in these flows has everywhere a positive Bernoulli constant.

While the very existence of strong outflows from small efficiency accretion black hole flows seems to be supported by observations, and therefore is most probably true, the reasons given by Blandford \& Begelman to explain the outflows, are certainly not correct, because, as discussed in \cite{AbrLasIgu20} by Abramowicz, Lasota \& Igumenshchev, (1)$\,$A positive Bernoulli constant is only a necessary, but certainly not a sufficient, condition for outflows\footnote{The {\it ``a-positive-Bernoulli-implies-strong-outflows''} is a wide spread fallacy. Rather surprisingly, as it is plainly incompatible with the well-known classic Bondi solution, in which fluid has a positive Bernoulli constant everywhere, but experiences no outflows \cite{KinFraRai02}.}. (2)$\,$The very argument that low efficiency accretion has a positive Bernoulli constant everywhere is not correct itself. It follows from an {\it ad hoc} mathematical assumption (made in order to make the problem easier to solve) that inefficient accretion flows are self-similar. Real flows obey inner and outer boundary conditions and for this reason are not self-similar. The boundary conditions imply {\it negative} Bernoulli constant at least close to both boundaries. 

\section{Summary and conclusions}

Observations provide a clear evidence for super-Eddington luminosities powered by accretion onto black holes. Theory, solidly based on standard physics, predicts that such super-Eddington luminosities imply highly super-Eddington accretion rates and a very low efficiency of the black hole accretion. The powerful and simple So{\l}tan argument shows, however, that the efficiency cannot be low all the time, because in the case of AGN this would be in a direct conflict with observations. 

Thus, observations and theory together seem to point that highly super-Eddington accretion, that really occur in several black hole sources, is a {\it transient} ``flip-flop'' phenomenon. A physical mechanism (or mechanisms ?) for the flip-flop behavior, in which periods of highly supper-Eddington but low efficiency accretion alternate with periods of highly efficient, sub-Eddington or nearly-Eddington accretion is not yet known. Its understanding presents a great challenge for all of us.

Neither ``adios'', nor any other theoretical model, could at present explain strong outflows and jets that are observed in several super-Eddington black hole sources. Understanding of such outflows is another great challenge. 

In the context of the cosmic structure formation, one should conclude that there is no obvious reason for assuming that ${\dot M}_{\rm Edd}$ is the upper limit for the growth rate of the seed black holes. 

\noindent {\it Acknowledgments:} I am grateful for suggestions, advice and help I received when writing this article from J{\'{\i}}rka Horak, Igor Igumenshchev, Jean-Pierre Lasota, William Lee, Piero Madau, Paola Rebusco, Martin Rees, Marek Sikora, Rashid Sunyaev and Andrzej Kajetan Wr{\'o}blewski.

%

\end{document}